\documentclass[conference,a4paper]{IEEEtran}
\usepackage[utf8]{inputenc}
\usepackage{graphicx}
\usepackage{tabularx}
\usepackage{amsmath}
\usepackage{amssymb}
\usepackage{stackengine}
\usepackage{makecell}
\usepackage{multirow}
\usepackage{cite}
\usepackage{array}
\usepackage{bbding}
\usepackage{pifont}
\usepackage{wasysym}
\usepackage{comment}
\usepackage{caption}
\usepackage{subcaption}
\usepackage[noend]{algpseudocode}
\usepackage{booktabs}
\usepackage{mathtools}
\usepackage{dirtytalk}
\usepackage{etoolbox}

\usepackage{algorithm,algpseudocode}

\title{
Crossover-BPSO Driven Multi-Agent Technology for Managing Local Energy Systems \vspace{-0.5cm}}
\makeatletter
\newcommand{\linebreakand}{%
  \end{@IEEEauthorhalign}
  \hfill\mbox{}\par
  \mbox{}\hfill\begin{@IEEEauthorhalign} \vspace{-10cm}
}
\makeatother
\author{
  \IEEEauthorblockN{Hafiz Majid Hussain}
  \IEEEauthorblockA{
       \textit{LUT University,}\\
    Lappeenranta 53850, Finland \\
    Majid.Hussain@lut.fi}
  \and
  \IEEEauthorblockN{Ashfaq Ahmad}
  \IEEEauthorblockA{
   \textit{Air University,}\\ 
    Islamabad 44000, Pakistan \\
    ashfaqahmad@ieee.org }
  \and
  \IEEEauthorblockN{Pedro H. J. Nardelli}
  \IEEEauthorblockA{
    \textit{LUT University,}\\
      Lappeenranta 53850, Finland \\
    Pedro.Nardelli@lut.fi} 
} 
\makeatletter
\patchcmd{\@maketitle}
  {\addvspace{0.5\baselineskip}\egroup}
  {\addvspace{-1\baselineskip}\egroup}
  {}
  {}
\makeatother

\begin{document}  
\IEEEoverridecommandlockouts
\IEEEpubid{\makebox[\columnwidth]{979-8-3503-9042-1/24/\$31.00~\copyright2024 IEEE\hfill} \hspace{\columnsep}\makebox[\columnwidth]{ }}
\maketitle
\IEEEpubidadjcol
\renewcommand\IEEEauthorblockA{\IEEEauthorblockA\vspace{-8\baselineskip}}

\begin{abstract}
This article presents a new hybrid algorithm, crossover binary particle swarm optimization (crBPSO), for allocating resources in local energy systems via multi-agent (MA) technology. Initially, a hierarchical MA-based architecture in a grid-connected local energy setup is presented. In this architecture, task specific agents operate in a master-slave manner. Where, the master runs a well-formulated optimization routine aiming at minimizing costs of energy procurement, battery degradation, and load scheduling delay. The slaves update the master on their current status and receive optimal action plans accordingly. Simulation results demonstrate that the proposed algorithm outperforms selected existing ones by 21\% in terms average energy system costs while satisfying customers' energy demand and maintaining the required quality of service.
\end{abstract}
\IEEEpeerreviewmaketitle

\vspace{-0.2cm}

\section{Introduction}
Local energy systems (LESs) are gradually transitioning from centralized and high-emission systems to hybrid and eco-friendly ones, mainly via DER (distributed energy resources) integration such as renewable energy sources (RESs), energy storage units, and demand response (DR) technologies \cite{stennikov2022coordinated}, \cite{hussain2020energy}. However, these integrations may face several challenges like the intermittency in RESs output power generation, supply uncertainty, and dynamic power system states posing stability implications \cite{yu2016mas}. In order to address these challenges, a multi-agent system (MAS) technology can be adopted for managing complex DER interactions within LES through a hierarchy of smart agents operating either independently or in collaboration to achieve either a common objective (\emph{e.g.}, resource optimization) or a set of objectives (like energy efficiency, grid stability, customer satisfaction, grid resilience). Hence, an advanced MAS technology driven by intelligent management and control is essential for ensuring green and cost-effective LES solutions  \cite{gonzalez2018multi}.

Previously, various energy management approaches have been explored to solve energy management problems in MAS to achieve economical solutions for energy users in LESs. For instance, authors in \cite{zadeh2023iot} employed MAS technology to reduce LES's operational expenses using a rule-based real-time corrective algorithm and DR strategy. The articles \cite{zhang2023multi} and \cite{alrobaian2023multi} presented MAS-based residential EMS (energy management system) to achieve multiple objectives including energy optimization, user satisfaction cost, and carbon emission reduction using distributed optimization and a greedy randomized adaptive search procedure. The authors in \cite{chreim2023energy} proposed a novel MAS-based EMS for energy internet-of-things (EIoT) devices to accomplish minimal energy loss and cost reduction by employing Lagrangian multiplier method. Similarly, the authors in \cite{pinto2018multi} proposed an intelligent MAS-based EMS architecture for smart buildings to achieve energy optimization and cost minimization via machine learning approaches (k-nearest neighbors algorithm and support vector machines). Similarly, refs. \cite{yu2016mas} and \cite{dou2016mas} presented a multi-layer MAS-based architecture to enhance reliable coordination via communication among various MAs such as DERs,  efficient EMS, and cost minimization. In other similar works, EMS architectures were proposed to achieve specific objectives like optimization of energy usage via peak-to-average (PAR) reduction through heuristic algorithms in \cite{li2017multiobjective} and \cite{hussain2018efficient}, and reduction of average system cost in \cite{ahmad2019real} via Lyapunov optimization. From an optimization standpoint, the approaches in in \cite{zadeh2023iot, zhang2023multi, alrobaian2023multi, chreim2023energy} have shown good performance in particular problem contexts, including situations with predefined rules, large-scale scenarios, and constraint-based issues. Nevertheless, the lack of flexibility and adaptability poses significant challenges when employing these approaches. To address these, heuristic algorithms (like genetic algorithm (GA), binary particle swarm optimization (BPSO), differential evolution (DE) and harmony search (HS)) are more effective in rapidly exploring complex search spaces to find near-optimal solutions, and flexibly adapting to diverse problem instances \cite{hussain2018efficient, ahmad2019real, yang2010engineering}.
\begin{figure*} [h!]
    \centering
    \includegraphics[height=8cm, width=15.cm]{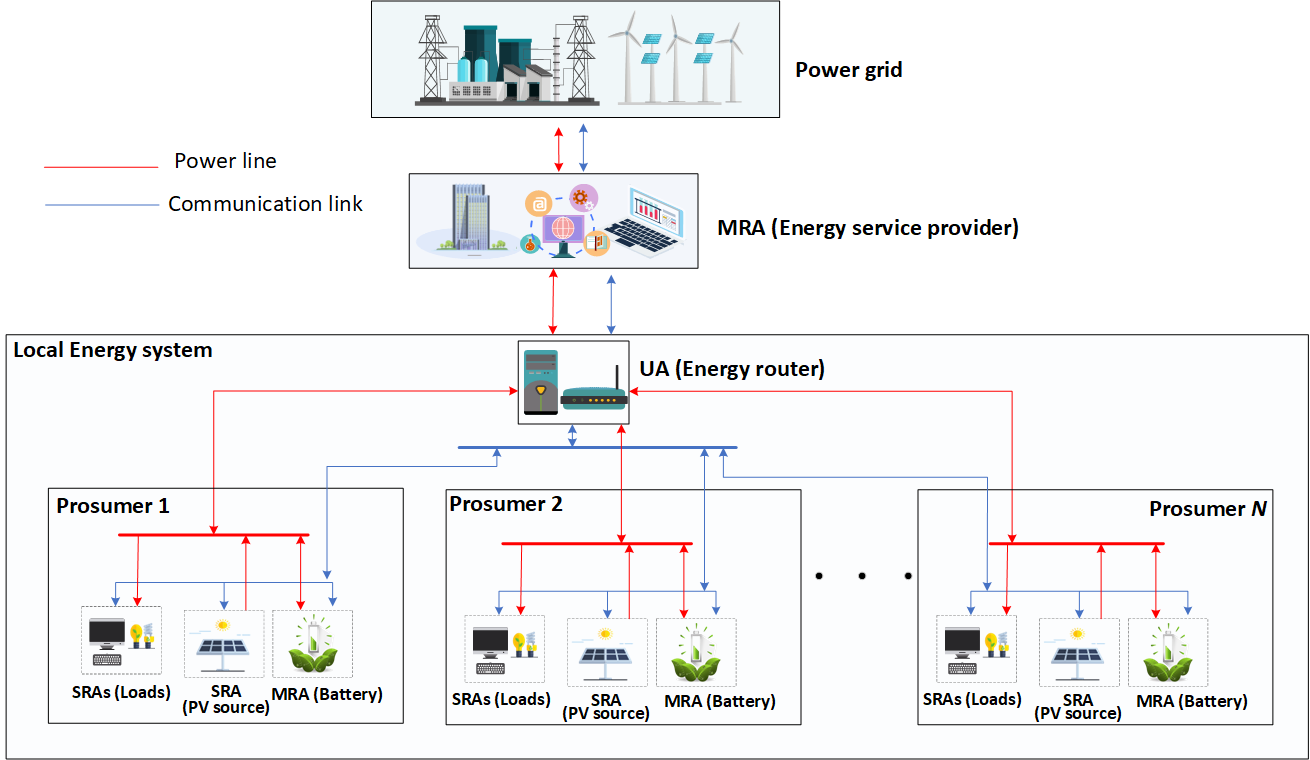} \vspace{-0.3cm}
    \caption{MA-based LES architecture involving SRAs (loads and PV source), MRAs (battery storage and ESP), UA (energy router) and the external power grid.}
    \label{fig_1} \vspace{-0.5cm}
\end{figure*}

Based on our comprehensive review of existing literature \cite{stennikov2022coordinated, yu2016mas, gonzalez2018multi, zadeh2023iot, zhang2023multi, alrobaian2023multi, chreim2023energy, pinto2018multi, dou2016mas, li2017multiobjective, hussain2018efficient, ahmad2019real}  above, it is evident that most prior studies have concentrated on exploring EMS solutions, either incorporating MAS technology or not. Despite their benefits, the following research gaps have been identified. \emph{First}: despite the diversity in EMS components and objectives, the agents designed and employed were homogeneous. For example, using either a group of simple reflex-based agents (SRAs) or model-based agents (MRAs) for maintaining a balanced energy load across a diverse set of devices and systems within a LES by dynamically adjusting electricity consumption in response to energy prices. This homogeneity limits the full potential of MAS technology, thereby reducing its effectiveness for EMS applications in LES. \emph{Second}: within the two main families of heuristic optimization methods, genetic evolution-based approaches lack elements of social learning, while swarm movement-based approaches lack genetic diversity. For example, the exploration phases of GA and DE involve diversity through a non-directed genetic recombination process, and that of PSO involve directed velocity-based position updates without taking into account recombination-based diversity \cite{yang2010engineering}. As a result, the exploratory phase in these algorithms is constrained, leading to convergence at local optima.
To address the aforementioned research gaps, this work introduces a hierarchical MA technology-based architecture powered by our newly proposed \emph{crBPSO} algorithm for EMS applications in LES. Then, simulations are carried to validate the relative effectiveness of the proposed work in terms of selected performance metrics. The main contributions and benefits of this research can be summarized as follows.

\begin{itemize}
    \item A novel hierarchical MA technology-based architecture is introduced, featuring heterogeneous agents (SRAs, MRAs and utility agent (UA)) designed and employed to account for diverse EMS components and objectives.
    \item  An energy transaction model is developed for energy trade between energy consumers and energy service provider (ESP). This model enables cost-effective solutions for energy consumers while maintaining demand-supply balance.
    \item A new hybrid heuristic algorithm, called crossover-BPSO (\emph{crBPSO}), is proposed. It integrates crossover from genetic algorithms into the velocity update equation of BPSO, enhancing exploration abilities and enabling deeper search space exploration.
\end{itemize}
\vspace{-0.1cm}
The rest of the paper is organized as follows. Section II presents the multi-gent system model, formulates the problem, and presents the \emph{crBPSO} algorithm. Section III discusses the simulation results, and section IV concludes the paper. References are provides at the end.

\vspace{-0.25cm}
\section{Multi-Agent System Model}
Consider a MA-based architecture shown in figure \ref{fig_1} that encompasses three types of agents: simple reflex agents (SRAs), model-based-reflex agents (MRAs), and a utility-based agent (UA). The upcoming sections provide details on the explicit models of MAs along with their hierarchical functioning to devise optimal energy packet transaction schedules between ESP and prosumers.

\vspace{-0.1cm}
\subsection{Simple-Reflex Agents}
In this work, smart home loads and PV source are modelled as SRAs, as both undertake rule-based decisions depending only on the current percept.

The smart energy community (\emph{i.e.}, LES) shown in figure 1 consists of an array of smart homes $b \in  \{1,2,3,\;...,\;B\}$ such that each smart home has multiple IoT-enabled loads $l \in \{1,2,3,.., L\}$ modelled as SRAs. The energy consumed by these agents is quantized in the form of energy packets at discrete time slots $t_s  \in  \{0, 1,2,3,...T_{s}-1\}$, such that, the total energy packet demand of homes in the entire LES over a time horizon $T_s$ is given by:
\begin{equation}
E_{T_{s}}^{B,L}=\sum_{b=1}^{B}\sum_{l=1}^{L}\sum_{t=0}^{T_{s}-1} \chi_{t}^{b,l}\times E_{t}^{b,l} \label{Eq1}
\end{equation}
where, $\chi_{t}^{b,l}$ represents a number energy packets required for smart home $b$ of load $i$ at time instant $t$. Note that during the scheduling process, a delay may arise which is computed as: $d_{ t}^{b,l}=\frac{\Psi_{x,t}^{b,l}-\Psi_{A,t}^{b,l}}{\lambda_{t, max}^{b,l}-\Psi_{O,t}^{b,l}}$. Here, $\Psi_{x,t}^{b,l}$, $\Psi_{A,t}^{b,l}$, $\lambda_{t, max}^{b,l}$, and $\Psi_{O,t}^{b,l}$ represent scheduling start time, length of operation, the maximum allowable delay, and energy departure time of load $l$ at $t$ in $b$, respectively. The average experienced delay for a load \say{$l$} of of home \say{b} during $T_s$ is calculated using \eqref{Eq2}. 
\begin{align}
& \overline{d}_{T_{s}}^{B,L}=\frac{1}{J}\sum_{b=1}^{B}\sum_{t=0}^{T_{s}-1}\sum_{l=L}^{B}d_{ t}^{b,l}, \label{Eq2} 
\end{align}
The upper and lower bounds on $\varphi_{t}^{b,l}$ are given in \eqref{Eq3}.
\begin{align}
& d_{t,min}^{b,l} \leq d_{t}^{b,l} \leq d_{t,max}^{b,l} \label{Eq3} 
\end{align}

The output energy of PV systems $\text{E}_{\text {PV}, t}$ depend on its conversion efficiency ($\eta_{pv}$), generator area ($A_{g}$), solar irradiance $I_{ir,t}$, temperature correction factor, and outdoor temperature ($O_{tem}$) and can be computed as $\text{E}_{\text {PV}, t}= \eta_{pv}\times A_{g}\times I_{ir,t} (1-0.005(O_{tem}((t)-25))$. Thus, the harvested PV energy of the LES over $T_s$ is given by:
\begin{equation}
 \overline{\text{E}}_{\text {PV}, {T_{s}}}^{B}= \frac{1}{J} \sum_{b=1}^{B}\sum_{t=0}^{T_{s}-1} \text{E}_{\text {PV}, t}^{b}  \label{Eq5}
\end{equation} 
It is assumed that the harvested energy is first given to serve scheduled loads at that time and the residue amount of $\text{E}_{\text {PV}, t}$ (if any) is used to charge the storage battery. 

\subsection{Model-based Reflex Agents}
In this paper, energy storage system (ESS) and ESP are modelled as MRAs, as both involve dynamic state transitions depending upon past occurrences.

Based on energy demand-and-supply conditions, the ESS can be either charged ($\alpha$), discharged ($\beta$) at time slot $t$. Let $\alpha^{(+)}_{t}=1$ indicate an ESS charging event such that $\alpha^{(+)}_{t} \triangleq \{\text{1,  if }\text{E}_{\text {PV}}^{b}+G_{u,t}>0; \text{0,  otherwise} \}$. Similarly, assume $\beta^{(-)}_{t}=1$ indicate an ESS discharging event such that $\beta^{(-)}_{t} \triangleq \{\text{1,  if }\beta_{t}^{b}>0; \text{0,  otherwise} \}$. Thereby, energy states of ESS evolve according to below given equations.
\begin{eqnarray}
& \text{ESS}_{t+1}^{b}= \vartheta_{t}\text{ESS}_{t}^{b}+ \eta_{t}^{(+)}\Big(\text{E}_{\text {PV}, t}^{j}+G_{u,t}^{g}\Big)-\eta_{t}^{(-)}\Big(\beta_{t}^{b}\Big)\label{Eq6} \\
& \text{ESS}_{t}^{b}=\alpha_{t}^{(+)}\kappa_{t}^{b,(+)}+\beta_{t}^{(-)}\kappa_{t}^{b,(-)} \label{Eq10}
\end{eqnarray}
In \eqref{Eq6}, $\vartheta_{t}, \eta_{t}^{(-)},$ and $\eta_{t}^{(+)}$ represent decay rate, charging efficiency and discharging  efficiency of ESS, respectively. Wherein \eqref{Eq10}, $\kappa (+)$ and  $\kappa (-)$  are the ESS degradation cost due to charging and discharging activities, respectively \cite{ahmad2019real}. From \eqref{Eq6} and \eqref{Eq10},  the average aggregated battery degradation cost can be computed as:
\begin{equation}
 \overline{\Pi}_{T_{s}}^{B}=\frac{1}{J}\sum\limits_{b=1}^{B}\sum\limits_{t=0}^{T_{s-1}}\text{ESS}_{t}^{b} \label{Eq11}
\end{equation}

Smart home prosumers are actively engaged in energy transactions with the ESP. On encountering energy deficit, they can buy energy packets from the main grid ($G_u$) via ESP. They can also sell back excess energy to the main grid via ESP. These energy transactions are carried with the consideration of constraints imposed by the feed-in-tariff policy of $G_u$ and the demand-and-supply ratio ($\text{R}_t^{ds}$) within the LES. The ESP buys energy packets from the power grid and smart home prosumers at prices $\Upsilon_{t}^{b, buy}$ and $\zeta_{t}^{G_u, buy}$, respectively. It can also sell back energy packets to the power grid and smart home prosumers at prices $\Upsilon_{t}^{b, sell}$ and $\zeta_{t}^{G_u, sell}$, respectively. Mathematically:
\begin{IEEEeqnarray}{rCl}
\Upsilon_{t}^{b, \text{buy}} &=& \Upsilon_{t}^{b, \text{sell}} (\text{R}_t^{\text{ds}})^2 H^2(1 - \text{R}_t^{\text{ds}}) + \zeta_{t}^{G_u, \text{sell}} (1 - H(\text{R}_t^{\text{ds}} - 1))  \nonumber \\
&& +\> \zeta_{t}^{G_u, \text{buy}} (1 - \text{R}_t^{\text{ds}}) H(\text{R}_t^{\text{ds}}) H(1 - \text{R}_t^{\text{ds}}) \IEEEyesnumber
\label{Eq12} \\
\Upsilon_{t}^{b, \text{sell}} &=& \frac{\zeta_{t}^{G_u, \text{sell}} \zeta_{t}^{G_u, \text{buy}}. H(\text{R}_t^{\text{ds}}) \cdot H(1 - \text{R}_t^{\text{ds}})}{(\zeta_{t}^{G_u, \text{buy}} - J \zeta_{t}^{G_u, \text{sell}}) \text{R}_t^{\text{ds}} + \zeta_{t}^{G_u, \text{sell}}} \nonumber \\
&& +\> \zeta_{t}^{G_u, \text{sell}} \cdot (1 - H(\text{R}_t^{\text{ds}} - 1)) \label{eq:MyEquation}
\end{IEEEeqnarray}
From \eqref{Eq12} and \eqref{eq:MyEquation}, energy can be procured or sold back to the grid based on $\text{R}^{ds}$: (i)  when $\text{R}^{ds}=0$, energy can be procured from the grid at  price $\zeta_{t}^{G_u, buy}$;  (ii) when $\text{R}^{ds}>=1$,  energy can be sold back to the grid at price $\zeta_{t}^{G_u, sell}$; and (iii) when $0<\text{R}^{ds} <1$, the selling price of energy dynamically fluctuates between  $\zeta_{t}^{G_u, sell}$ and $\zeta_{t}^{G_u, buy}$. Further, the costs of energy procured by prosumer from the grid and sold back to the grid are given by \eqref{Eq14} and \eqref{Eq15}, respectively. 
\begin{IEEEeqnarray} {rCl}
&& \Pi_{t}^{b,\text{buy}}=\Upsilon_{t}^{b, sell} \Big(E_{t}^{B,L}-(E_{PV,t}^{b}+\text{ESS}_{t}^{b})\Big)\;\;
\label{Eq14} \\
&& \Pi_{t}^{b,\text{sell}}=\Upsilon_{t}^{b, buy} \Big((E_{PV,t}^{b}+\text{ESS}_{t}^{b})-E_{t}^{B,L}\Big)\;\;
\label{Eq15}
\end{IEEEeqnarray}
From \eqref{Eq14} and \eqref{Eq15}, the average aggregated cost of the energy transactions over $T_s$ horizon is given in \eqref{Eq13} below.
\begin{equation}
\overline{\Pi}_{T_{s}}^{Bx}=\frac{1}{J}\sum_{t=0}^{T_{s}-1} \sum_{b=1}^{M}\Big( \Pi_{t}^{b,\text{sell}} -
\Pi_{t}^{b,\text{buy}}\Big).\label{Eq13}
\end{equation}

\subsection{Utility Agent} \label{SecUA}
This paper models energy router as a UA, as it aims to minimize a utility function which is an average aggregated system cost.

Let $\Phi \triangleq [G_u,\text{E}_{\text {PV}, t}^{k,b}, \text{ESS}_{t}^{b}, \text{E}_{\text {PV}, t}^{b}]$ be a vector of  control actions of the UA for prosumer $b$ at time slot $t$. The UA's objective is to minimize the average aggregated system cost, which encompasses three factors: (i) the cost associated with the scheduling delay of the loads $\Pi(\overline{d}_{T_{s}}^{B,L})$, (ii) the cost of energy packet transactions involving both buying and selling ($\overline{\Pi}_{T_{s}}^{Bx}$), and (iii) the cost of energy storage system degradation ($\overline{\Pi}_{T_{s}}^{B}$). The problem can be formulated as follows.
\begin{IEEEeqnarray}{rCl}
&& \text{\stackunder{Minimize }{$\{\Phi_{T_s}, {d}_{T_{s}}^{B,L}\}$}\quad} \Pi(\overline{d}_{T_{s}}^{B,L})+\overline{\Pi}_{T_{s}}^{Bx}+\overline{\Pi}_{T_{s}}^{B} \label{obj} \\
&& \text{Subject to:} \nonumber \\
&& 0 \leq\Upsilon_{t}^{b,buy} \leq \Upsilon_{t,max}^{b,buy}, \quad 0 \leq \Upsilon_{t}^{b,sell} \leq \Upsilon_{t,max}^{b,sell} \label{22} \\
&& \sum\limits_{b=1}^{B}\sum\limits_{l=1}^{L}\sum\limits_{t=0}^{T_{s}-1}Q_{t}^{b,l}=E_{T_{0}}^{B,L} \label{20} \\
&& E_{T_{s}}^{B,L}- Q_{t}^{b,l} \leq \mathcal{N}_{max} \label{24} \\    
&& \text{E}_{\text {PV}, t}^{k,b} = \text{min} \{R_{t}^{B,L},\text{E}_{\text {PV}, t}^{b} \} \label{25} \\
&& 0 \leq \text{E}_{\text {PV}, t}^{b} \leq \text{E}_{\text {PV}, t}^{b} -\text{E}_{\text {PV}, t}^{k,b} \label{26} 
\end{IEEEeqnarray}
\begin{IEEEeqnarray}{rCl}
&& \text{E}_{\text {PV}, t}^{b}+ G_u \in \Bigr[0, \text{min} \{\text{ESS}_{t,max}^{b}, \text{ESS}_{t,max}^{b}-\text{ESS}_{t}^{b}\}\Bigr] \label{27} \\
&& \beta_{t}^{k,b} \in \Bigr[0, \text{min} \{ \beta_{t,max}^{k,b},\text{ESS}_{t}^{b}-\text{ESS}_{t,min}^{b}\}\Bigr] \label{28} \\
&& 0 \leq \text{E}_{\text {PV}, t}^{b}+G_{u,t} \leq \text{ESS}_{t,max}^{b} \label{29} \\
&& \text{ESS}_{t,min}^{b} \leq   \text{ESS}_{t}^{b} \leq  \text{ESS}_{t,max}^{b} \label{31}  
\end{IEEEeqnarray}
In the above formulation, constraints \eqref{22}-\eqref{24} pertain to energy buying and selling. Constraint \eqref{20} indicates that energy packets can be scheduled within the time horizon $T_s$, while ensuring that total energy packet demands remain constant before and after scheduling. Additionally, constraint \eqref{24} limits the selling of excess generated energy ($\mathcal{N}_{max}$) to utilities for grid security reasons. Constraints \eqref{25}-\eqref{31} are related to $\text{E}_{\text {PV}, t}^{b}$ and $\text{ESS}_{t}^{b}$. Specifically, \eqref{25} and \eqref{26} indicate that $\text{E}_{\text {PV}, t}^{b}$ irst addresses the appliance requests in any $b$ at time slot $t$, and and any remaining energy ($\text{E}_{\text {PV}, t}^{k,b}>0$) can be stored in $\text{ESS}_{t}^{b}$. Constraints in \eqref{27} and \eqref{28} account for operations involving ESS. Finally, constraints \eqref{29} and \eqref{31} set the upper and lower bounds for charging and discharging activities, as well as battery capacity.
\subsubsection{\textbf{The crBPSO Algorithm}}
In order to achieve the objective function in \eqref{obj} subject to constraints \eqref{22}--\eqref{31}, a new algorithm is presented here. This algorithm adopts a hybrid approach by integrating the crossover operation of GA into the velocity update procedure of BPSO, thereby enhancing its exploration capabilities beyond those of both BPSO and GA. The specifics of the phase-wise algorithm are detailed below.
\begin{enumerate}
    \item \emph{Initialization:} set values of algorithm parameters $\{\alpha, {v_{t}}^{i}, v_{t,max}^{cr}, v_{t,min}^{cr}$\}, and employ a uniform random distribution to generate an initial population of candidate solutions such that each $i$-th particle has a position $s_{0}^{i}$ as per \eqref{A_1} given below:
\begin{equation}\label{A_1}
\begin{split}
s_{0}^{i} = & \left( x_{min} + r_{0,x}^{i}(x_{max}-x_{min}), \right. \\
           & \left. \quad y_{min} + r_{0,y}^{i}(y_{max}-y_{min}) \right) 
\end{split}
\end{equation}
          where, $r_{0,x}^{i}$ and $r_{0,y}^{i}$ are random numbers between 0 and 1, both following a uniform distribution. And $x_{min}$, $x_{max}$, $y_{min}$ and $y_{max}$ are lower and upper bounds on the x-and-y coordinates, respectively.
          
    \item \emph{Evaluation:} evaluate the generated initial population $s_{0}^{i}$ in (23) based on the objective function in (13), and the constraints in (14)--(22).
    \item \emph{Exploration}: BPSO algorithm explores the search space (positions of particles) in (\ref{PSO2}) with the help of their velocities in (\ref{PSO1}), both adopted from \cite{del2008particle}. 
        \begin{align}
            &  v_{t}^{i}= v_{t-1}^{i}+ \alpha_{1}r_{1}(s_{r}-s_{t-1}^{i}) +  \alpha_{2}r_{2}(s_{g}-s_{t-1}^{i}) \label{PSO1} \\
            & s_{t}^{i}= s_{t-1}^{i}+v_{t}^{i}   \label{PSO2}
        \end{align}
        Where, ${v_{t-1}}^{i}$, ${v_{t}}^{i}$, ${s_{t-1}}^{i}$, $s_{t}^{i}$, and ${s_{r}}^{i}$ indicate previous velocity, current velocity, previous position, current position, and local best position of each $i$-th particle, respectively. And, ${s_{g}}$ is the symbol for the global best position of the swarm. $r_1$ and $r_2$ in \eqref{PSO1} are random number between 0 and 1 both following a uniform distribution, $\alpha_1$ is coefficient of cognitive learning and $\alpha_2$ is coefficient of social learning. The BPSO algorithm, which utilizes the directed search method based on equations \eqref{PSO1} and \eqref{PSO2}, has limitations that affect its effectiveness. Notably, the particles or individuals within the BPSO algorithm tend to favor the local best solution due to the formation of a Markov chain. This tendency towards a dominant local solution is further reinforced through deterministic attraction motions observed in the population during multiple Markov chain processes. As a result, the exploratory phase within the BPSO algorithm is constrained, leading to limited exploration capabilities \cite{yang2010engineering, del2008particle}. Note that GA demonstrates potential exploration abilities, however, its exploration phase is constrained in terms of directed search. To address these limitations, the \emph{crBPSO} algorithm incorporates crossover (borrowed from GA) into the velocity update equation of BPSO. This integration enhances the exploration abilities of the algorithm, allowing it to delve deeper into the search space. The crossover operation is implemented through \eqref{A_3} and \eqref{CR}.
            \begin{align}
            &  v_{t}^{i}= v_{t-1}^{i}+ \alpha v_{t}^{cr} \label{A_3} \\
            & v_{t}^{cr} = \begin{cases}
                                r_{1}(s_{r}-s_{t-1}^{i}) & \text{{with probability}} \hspace{2pt} (P_{cr}) \\
                                 r_{2}(s_{g}-s_{t-1}^{i})& \text{with probability}(1-P_{cr})
                            \end{cases} \label{CR} \vspace{-0.6cm}
        \end{align} 
        It is now evident that equations \eqref{A_3} and \eqref{CR} provide the \emph{crBPSO} with enhanced effectiveness compared to GA and BPSO. As a result, the proposed \emph{crBPSO} ensures an efficient exploration process by thoroughly covering promising regions of the search space.
        \item \emph{Re-evaluation}: re-evaluate the updated population $s_{t}^{i}$ in step 3 above based on the objective function in (13), and the constraints in (14)--(22). Stop if convergence is achieved, otherwise go back to step 3 above.
\end{enumerate}
\section{Results and Discussions}
\begin{figure}[h!]
\includegraphics[width=1\linewidth]{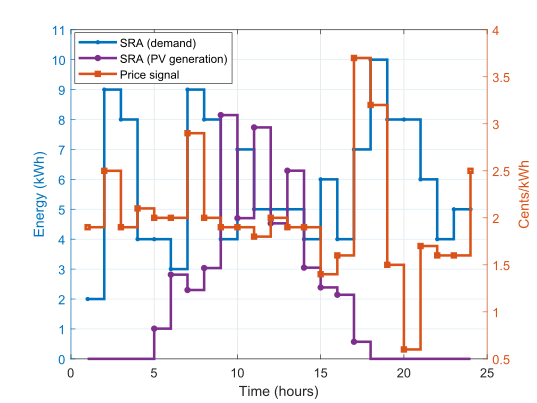} \vspace{-0.6cm}
    \caption{Profiles of energy demand (SRA), PV generation (SRA) and price signal (MRA).}
\label{Profiles} \vspace{-0.6cm}
\end{figure}

In simulations, we consider a MA-based LES with multiple smart home prosumers (J=10) having identical energy profiles over a 24-hour period. Profiles of energy demand (SRA), PV generation (SRA) and price signal (MRA) are shown in figure \ref{Profiles}. This figure shows that prosumers engage in energy transactions with the power grid via an ESP, with buying prices ranging from $\Pi_{t,min }^{b,\text{buy}}=1.5$ cents/kWh to $\Pi_{t, max }^{b,\text{buy}}=9.5$ cents/kWh. The maximum average power output ($\overline{\text{E}}_{\text {PV}, {t, max}}^{b}$ ) is 2.85 kWh, with a PV efficiency ($\eta{pv}$) of 18\%. The ESS parameters are set as follows: a capacity of $\overline{\text{ESS}}_{t, max}^{b} = 5$ kWh, a self-discharge rate of $\vartheta_t = 0.8$, and charge/discharge efficiencies of $n_t^{(+)} = n_t^{(-)} = 0.7$.

\begin{figure}[h!]
\includegraphics[width=1\linewidth]{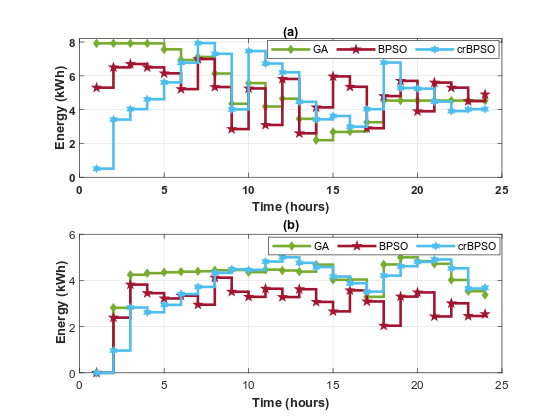} \vspace{-0.7cm}
    \caption{Optimal energy allocation plans of the MA-based architecture under GA, BPSO and \emph{crBPSO} algorithms: (a) for SRAs, and (b) for MRAs.}
\label{P_E} \vspace{-0.5cm}
\end{figure}

As mentioned in section \ref{SecUA}, the UA runs \emph{crBPSO}  algorithm to devise reduced cost-based energy plans for SRAs and MRAs at prosumer premises in the LES. In this regard, comparative results of \emph{crBPSO} with both GA and BPSO are shown in figure  \ref{P_E}(a) and \ref{P_E}(b). It is evident from the results that energy demands of SRAs and MRAs are scheduled efficiently by UA through a combination of on-site renewable sources and power supplied from the external grid. Although the PV energy production does not fully meet the energy needs of SRAs, the optimization algorithms exploit low, mid, and high price periods very well in devising energy plans for the considered agents. Notably, the \emph{crBPSO} algorithm schedules most of the energy demand in low and mid peak periods compared to both GA and BPSO, hence it is relatively more efficient in devising energy plans for smart home prosumers. This is also evident from the average daily and monthly prosumers' energy bills in table \ref{Table:buy_sell}. \vspace{-0.5cm}
\begin{table}[h!]
\caption{Energy transactions between UA and MRAs.}
\renewcommand{\arraystretch}{1}
\setlength{\tabcolsep}{5pt}
\label{Table:buy_sell} 
\begin{center}
\begin{tabular}  {ccccc}
\toprule
&{${\overline{\Upsilon}_{T_s}^{B,buy}}$ } & {${\overline{\Upsilon}_{T_s}^{B,sell}} $ }&\thead{Daily \\bill (\$) }&\thead{Monthly \\bill (\$) }
\\\hline
{Unscheduled}   & 1.03  &0.73 & 0.3&9\\\hline
{GA} &0.74 &0.87  &0.13&3.9\\  \hline
\hline
{BPSO}  &1.38   &1.70 &-0.32   & -6.9\\\hline
{\emph{crBPSO}}    &1.03 &1.07  &-0.04   & -1.2\\\hline
\bottomrule
\vspace{-1cm}
\end{tabular}
\end{center}
\end{table}

\section{Conclusion}
This article introduced a new hybrid algorithm, called \emph{crBPSO}, designed to optimize the allocation of energy resources within a LES using a MA framework. The system employed a hierarchical MA architecture in a grid-connected LES setup, structured in a master-slave fashion. The master agent was responsible for resource allocation among slave agents, formulated as a joint scheduling optimization problem. To solve this, the \emph{crBPSO} algorithm was proposed, aiming to minimize the average aggregated energy system cost, encompassing energy procurement, battery degradation, and scheduling delays. Comparative simulations reveal that the \emph{crBPSO} algorithm significantly reduced the average aggregated energy system cost and provided efficient energy plans for smart home prosumers.
\vspace{-0.3cm}
\section*{Acknowledgements}
\vspace{-1ex}
This paper was partly funded by EU MSCA project COALESCE (n.101130739), and by Research Council of Finland via X-SDEN (n.349965), EnergyNet (n.321265/n.328869/n.352654) and ECO-NEWS (n.358928).

\bibliographystyle{ieeetr}
\bibliography{Ref}
\end{document}